\documentclass[a4paper, amsfonts, amssymb, amsmath, reprint, showkeys, nofootinbib, oneside]{revtex4-2}%onecolumn,
\usepackage[left=2cm, right=2cm, top=2cm, bottom=2cm]{geometry}
\usepackage{graphicx}% Include figure files

\usepackage[pdftex, pdftitle={Article}, pdfauthor={Author}]{hyperref} % For hyperlinks in the PDF
\usepackage[normal]{caption2}
\usepackage[utf8]{inputenc}  % 支持UTF-8编码
\usepackage[T1]{fontenc}     % 优化字符输出
\usepackage[english]{babel}
\usepackage{ragged2e} % 导言区

\begin{document}

\preprint{APS/123-QED}

\title{Physics-informed neural network enhanced multispectral single-pixel imaging with a chip spectral sensor}% Force line breaks with \\

\author{Muchen Zhu\textsuperscript{1}}
\author{Baolei Liu\textsuperscript{2}}
 \email{liubaolei@buaa.edu.cn}
\author{Yao Wang\textsuperscript{1}}
\author{Linjun Zhai\textsuperscript{1}}
\author{Jiaqi Song\textsuperscript{1}}
\author{\\Nana Liu\textsuperscript{3,4}}
\author{Zhaohua Yang\textsuperscript{2}}
\author{Lei Ding\textsuperscript{5}}
\author{Fan Wang\textsuperscript{1}}

\affiliation{%
 \textsuperscript{1}School of Physics, Beihang University, Beijing 102206, China\\    
 \textsuperscript{2}School of Instrumentation and Optoelectronics Engineering, Beihang University, Beijing 100191, China\\    
 \textsuperscript{3}College of Materials Science and Chemical Engineering, Harbin Engineering University, Harbin, 150001, Heilongjiang, China\\
\textsuperscript{4}Qingdao Innovation and Development Center of Harbin Engineering University, Qingdao 266500, Shandong, China\\
\textsuperscript{5}Centre for Atomaterials and Nanomanufacturing, School of Science, RMIT University, Melbourne, Victoria, Australia
}%

\date{Augest 9, 2025}% It is always \today, today,

\begin{abstract}
Multispectral imaging (MSI) captures data across multiple spectral bands, offering enhanced informational depth compared to standard RGB imaging and benefiting diverse fields such as agriculture, medical diagnostics, and industrial inspection. Conventional MSI systems, however, suffer from high cost, complexity, and limited performance in low-light conditions. Moreover, data-driven MSI methods depend heavily on large, labeled training datasets and struggle with generalization. In this work, we present a portable multispectral single-pixel imaging (MS-SPI) method that integrates a chip-sized multispectral sensor for system miniaturization and leverages an untrained physics-informed neural network (PINN) to reconstruct high-quality spectral images without the need for labeled training data. The physics-informed structure of the network enables the self-corrected reconstruction of multispectral images directly with the input of raw measurements from the multispectral sensor. Our proof-of-concept prototype achieves the reconstruction of 12-channel high-quality spectral images at the sampling rate of 10\%. We also experimentally validate its performance under varying sampling rate conditions, by comparing it with conventional compressive sensing algorithms. Furthermore, we demonstrate the application of this technique to an MSI–based image segmentation task, in which spatial regions are discriminated according to their characteristic spectral signatures. This compact, high-fidelity, and portable approach offers promising pathways to lightweight and cost-effective spectral imaging on mobile platforms.
\end{abstract}

%\keywords{Computational spectrometer, on-chip, electrochromic, spectral resolution}
%Use show keys class option if keyword
\maketitle

%\tableofcontents

\section{Introduction}
Multispectral imaging (MSI) has found widespread applications across agriculture \cite{r1}, medical diagnostics \cite{r2}, industrial inspection \cite{r3}, environmental monitoring \cite{r4}, and astronomy \cite{r5}. These systems acquire images across multiple discrete spectral bands, providing richer information than standard RGB cameras. With the development of compact lens arrays \cite{r6}, snapshot MSI sensors \cite{r7,r8}, and computational imaging techniques \cite{r9}, MSI technologies have advanced significantly over the past decade in terms of spatial resolution, spectral coverage, and acquisition speed. However, conventional MSI techniques still have several limitations, such as high cost and high complexity of imaging systems, limited signal-to-noise ratio under low-light conditions, and heavy data storage demands. These issues hinder the adoption of MSI technologies in portable applications and consumer devices.

\begin{figure*}[htbp]
%\centering
\includegraphics[width=0.9\textwidth]{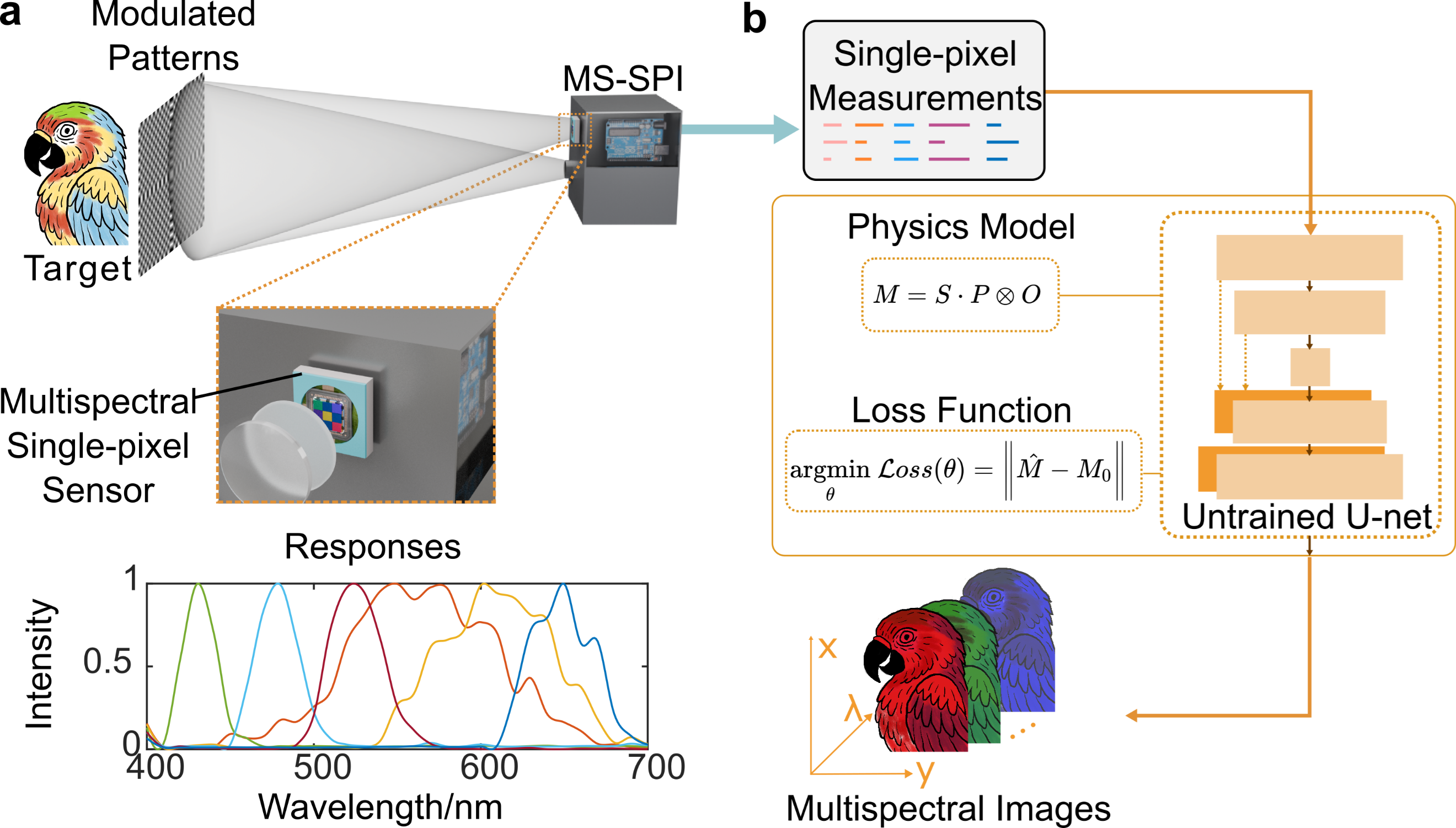}
\caption{\justifying Framework of the MS-SPI system. (a) Binary spatial modulated patterns are projected onto the scene. A chip-sized multi-spectral sensor, which serves as multiple single-pixel detectors at different wavebands, is adopted to measure the reflected light intensities from the target. (b) The raw single-pixel measurements are input to the physics-informed neural network (PINN) to reconstruct the 3D multispectral data cube.}\label{fig:1}
\end{figure*}

Single-pixel imaging (SPI), as a rapidly developing computational imaging technique, emerges as a promising alternative for conventional camera-based imaging techniques \cite{r10}. SPI reconstructs spatial information from a series of time-varying single-pixel light intensities captured by a non-spatially resolving detector, unlike traditional imaging methods that rely on pixelated detector arrays \cite{r11,r12}. SPI enables high-fidelity image reconstruction with fewer measurements than the pixel numbers via compressive sensing (CS), thereby reducing data acquisition time and system complexity. In the past decades, the spectral response range of SPI has been expanded from visible light to UV \cite{r13}, infrared \cite{r14}, and even x-ray \cite{r15} or terahertz wavelengths \cite{r16}. SPI based spectral imaging can also be achieved with simplified hardware implementations, compared with conventional MSI systems \cite{r17}. Recent advances in multispectral single-pixel imaging (MS-SPI) include the use of multiple single-pixel detectors \cite{r18}, spectral-encoded illumination patterns \cite{r19}, and frequency multiplexed illumination \cite{r20}. For example, MS-SPI technology achieves low-cost and wide-spectrum multispectral imaging through Fourier decomposition on temporal signals with a single-pixel detector \cite{r21}. By modulating spatial information and dispersive spectra in regions on the DMD, the MS-SPI system achieves high-throughput spectral video recording at low bandwidth \cite{r22}. The cascaded compressed-sensing single-pixel cameras achieve hyperspectral imaging with a spectral resolution of 6.2 nm \cite{r23}. These advances highlight the achievements of single-pixel methods for enabling flexible and cost-effective spectral imaging. Moreover, the integration of deep neural networks and MS-SPI further enhances the quality of reconstructed images from under-sampled measurements \cite{r24,r25,r26,r27}. However, data-driven neural network approaches remain limited by their dependence on large-scale, labeled datasets \cite{r28}, which are acquired costly and labor-intensive. Moreover, data-driven networks generalize poorly to unseen scenes or spectral distributions due to a lack of physical constraints, which reduces their reliability.

In this work, we present a physics-informed neural network (PINN) enhanced portable MS-SPI method that integrates with a chip-sized spectral sensor, for miniaturized hardware implementation and high-quality reconstruction without pre-training. We utilize a compact digital projector and the multispectral chip sensor to acquire spatially modulated light intensities across multiple spectral bands. By embedding the physical model of MS-SPI into the reconstruction neural network, we enable physics-corrected reconstruction of spectral images without the annotated dataset. Our prototype system achieves 12-channel high-quality spectral image reconstruction at a sampling rate as low as 10\%. We also experimentally evaluate the performance of the reconstruction method under varying sampling rates and iteration steps. Furthermore, we demonstrate an example application of our approach with a proof-of-concept multispectral-assisted image segmentation, where the reconstructed spectral images are used for downstream analysis tasks. PINN-based MS-SPI framework provides a reliable, compact, and economical solution for efficient multispectral imaging, combining the interpretability of physical modeling with the flexibility of deep learning. This approach opens up possibilities for portable spectral analysis in resource-constrained scenarios, delivering lightweight construction and cost-performance efficiency for practical applications.
\section{Results}

The MS-SPI system is illustrated in Figure \ref{fig:1}. Figure \ref{fig:1}(a) shows the hardware setup. The device projects the spatially modulated binary patterns onto the target. Then we use a multispectral single-pixel sensor, which contains multiple single-pixel detectors with individual spectral filters (bottom part), to measure the reflected light intensities from the target. The single-pixel measurements obtained from the detectors are processed through a physics-informed neural network to reconstruct a 3D spectral data cube, as shown in Figure \ref{fig:1}(b). For the network, here we adapt an untrained U-net in the architecture, which is initialized randomly and optimized solely based on measurements \cite{r29}. We ensure a physics-corrected recovery process of the spectral data by the physics model that describes the MS-SPI in the reconstruction pipeline, where the loss function minimizes the error between the predicted measurements and the raw signals. This allows for efficient reconstruction even in the absence of large-scale, labeled datasets, making the network highly adaptable and efficient for MS-SPI.

\begin{figure*}[htbp]%% placement specifier
    \centering%% For centre alignment of image.
    \includegraphics[width=0.9\textwidth]{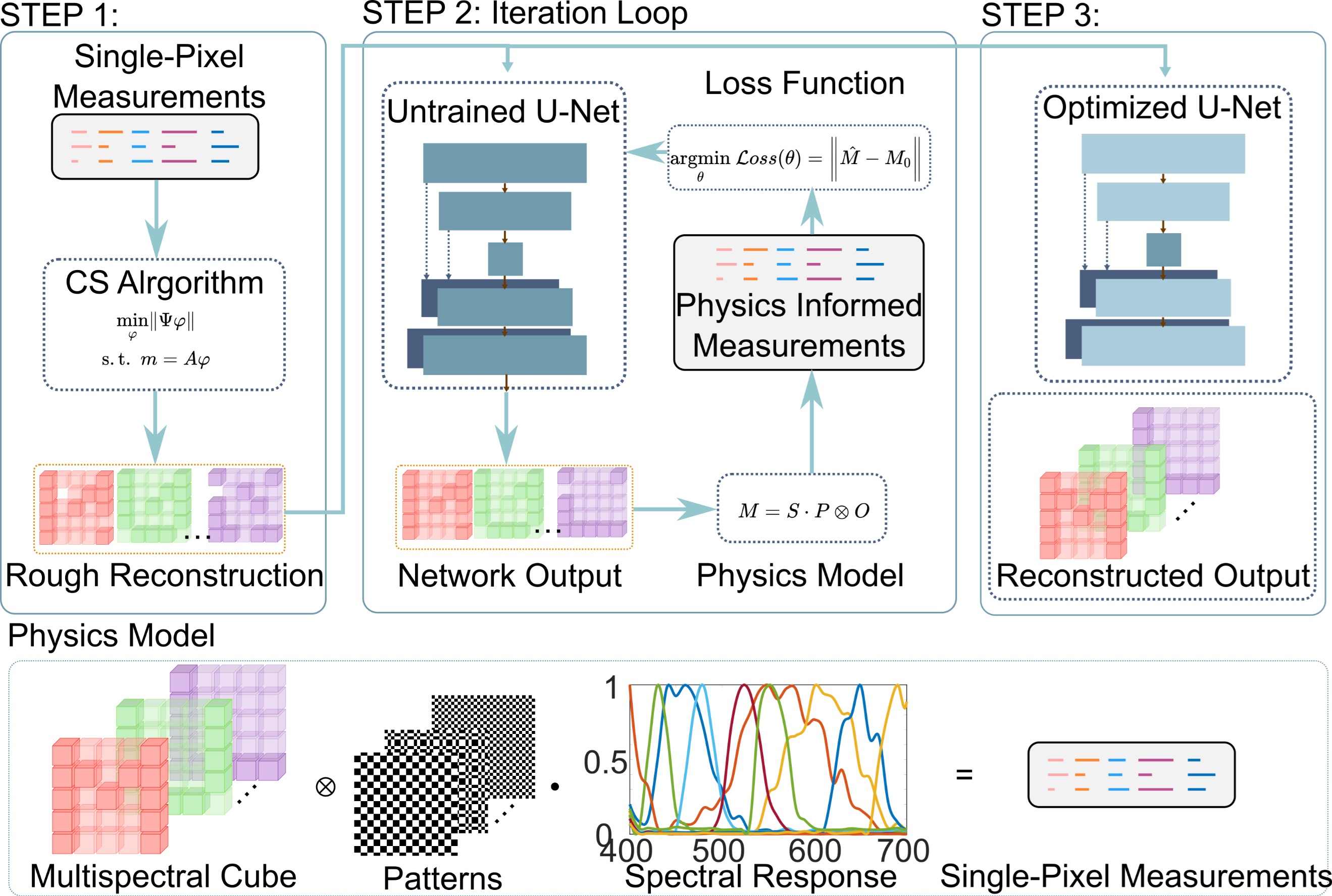}
    %% Use \caption command for figure caption and label.
    \caption{\justifying Architecture of the physics-informed reconstruction pipeline. The raw measurements are first used to reconstruct a rough image by using a CS solver (e.g., TVAL3). The result in Step 1 is then refined via an untrained U-Net, which is optimized through iterations using a physics-based forward model, as shown in the bottom, and a loss function defined on measurement fidelity. The U-Net output stepwise converges to an accurate multispectral reconstruction.}\label{fig:2}
\end{figure*}

Due to the small sensing area and limited light throughput of the miniature sensor, we employ a dedicated physics-informed reconstruction pipeline to ensure high-quality multispectral imaging. As illustrated in the top part of Figure \ref{fig:2}, the reconstruction framework consists of three main stages. First, we use the CS algorithm (TVAL3 \cite{r30}) to reconstruct a rough multispectral cube $\widetilde{O}$ from raw measurement $M$. Subsequently, the untrained U-net takes the rough reconstruction as input and produces an optimized multispectral estimate. This output is then passed through a forward physical model to generate predicted measurements$\hat{M}$. The Adam optimizer updates the network parameters by minimizing the discrepancy between the predicted and measured data, based on the loss function, where theta represents the learnable parameters of the network:

\begin{equation}
\begin{matrix}\underset{\theta }{\mathit{argmin}}\mathit{Loss}\left(\theta \right)=\left\|\widehat 
M-M\right\|.\end{matrix}
\end{equation}

Through iterative optimization, the predicted measurements produced by the network gradually converge to the sensor observations. In the final stage, the optimized U-Net serves as the final reconstruction module, transforming the initial estimate into a physically consistent and high-fidelity multispectral reconstruction.

In addition, the physics model of MS-SPI is shown in the bottom part of Figure \ref{fig:2}. The three-dimensional data cube of multispectral images comprises the $x$-axis, $y$-axis in space, and $\lambda$-axis in spectrum. Binary illumination patterns spatially modulate the original data cube, and the sensor measures the reflected light intensities. Concurrently, in the spectral domain, the multispectral sensor applies wavelength-dependent spectral responses, modulating the detected signal according to its sensitivity at each wavelength. The measurement $M_{m,n}$ corresponding to the $m$-th spatial pattern and $n$-th spectral channel can be expressed as:

\begin{equation}
\begin{matrix}$$M_{m,n}=\sum \limits_{x_i,y_j,\lambda _k}S_n\left(\lambda _k\right)P_m\left(x_i,y_j\right)O\left(x_i,y_j,\lambda
_k\right),$$\end{matrix}
\end{equation}where $M_{m,n}$ is the intensity recorded by the sensor, $P_m$ is the $m$-th projected spatial pattern, and $S_n$ is the spectral response function for the n-th channel. Given$ M$ spatial patterns and N spectral channels, a total of $M\times N$ measurements are acquired to reconstruct the data cube $O(x_i,y_j,\lambda_k)\in\mathbb{R}^{I\times J\times K}$, where $I,J$ denote spatial resolution and $K$ denotes the number of spectral bands. In general, by providing the representation of the MS-SPI measurement process, the proposed model supports the deployment of compact, chip-sized sensors while ensuring high-quality reconstruction through PINN.

\begin{figure}[htbp]%% placement specifier
    %\centering%% For centre alignment of image.
    \centering
    \includegraphics[width=\columnwidth]{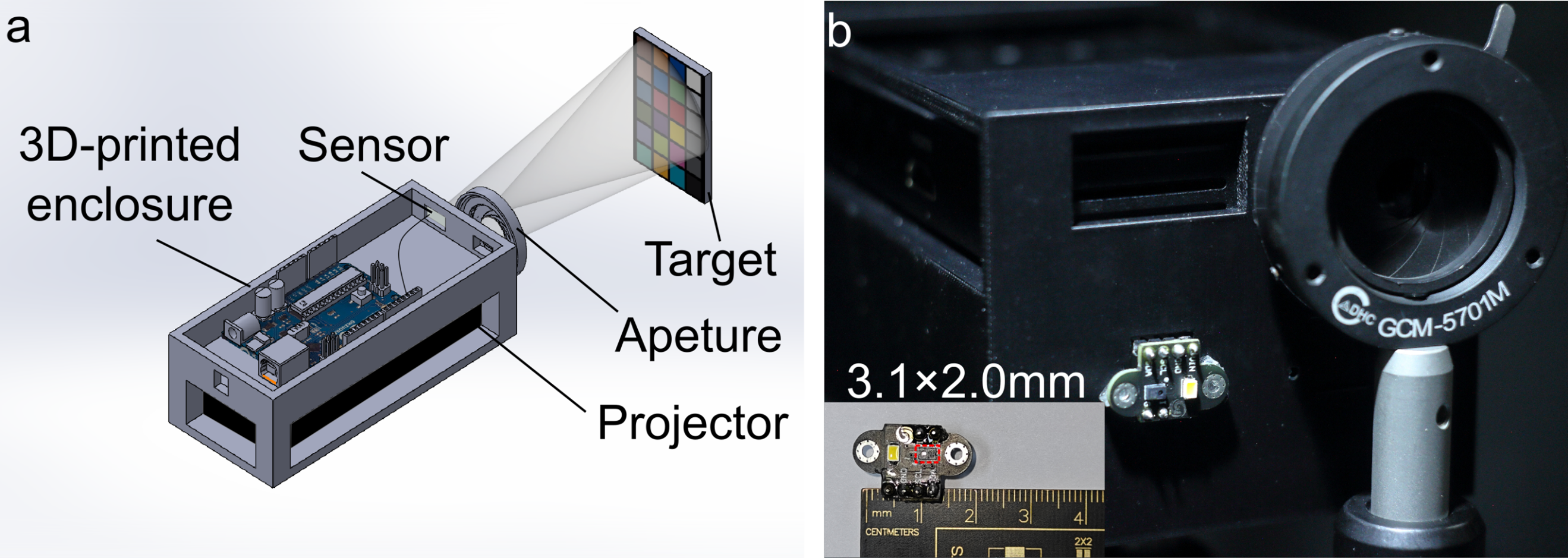}
    %% Use \caption command for figure caption and label.
    \caption{\justifying Experimental setup of the portable MS-SPI system. (a) Schematic diagram of the hardware setup, consisting of a minimized projector, a multispectral sensor, and an aperture, all integrated into a 3D-printed enclosure. (b) Photograph of the MS-SPI system (viewed from the front). The inset image shows the photograph of the multispectral sensor.}\label{fig:3}
\end{figure}
\begin{figure*}[htbp]%% placement specifier
\centering%% For centre alignment of image.
\includegraphics[width=0.8\textwidth]{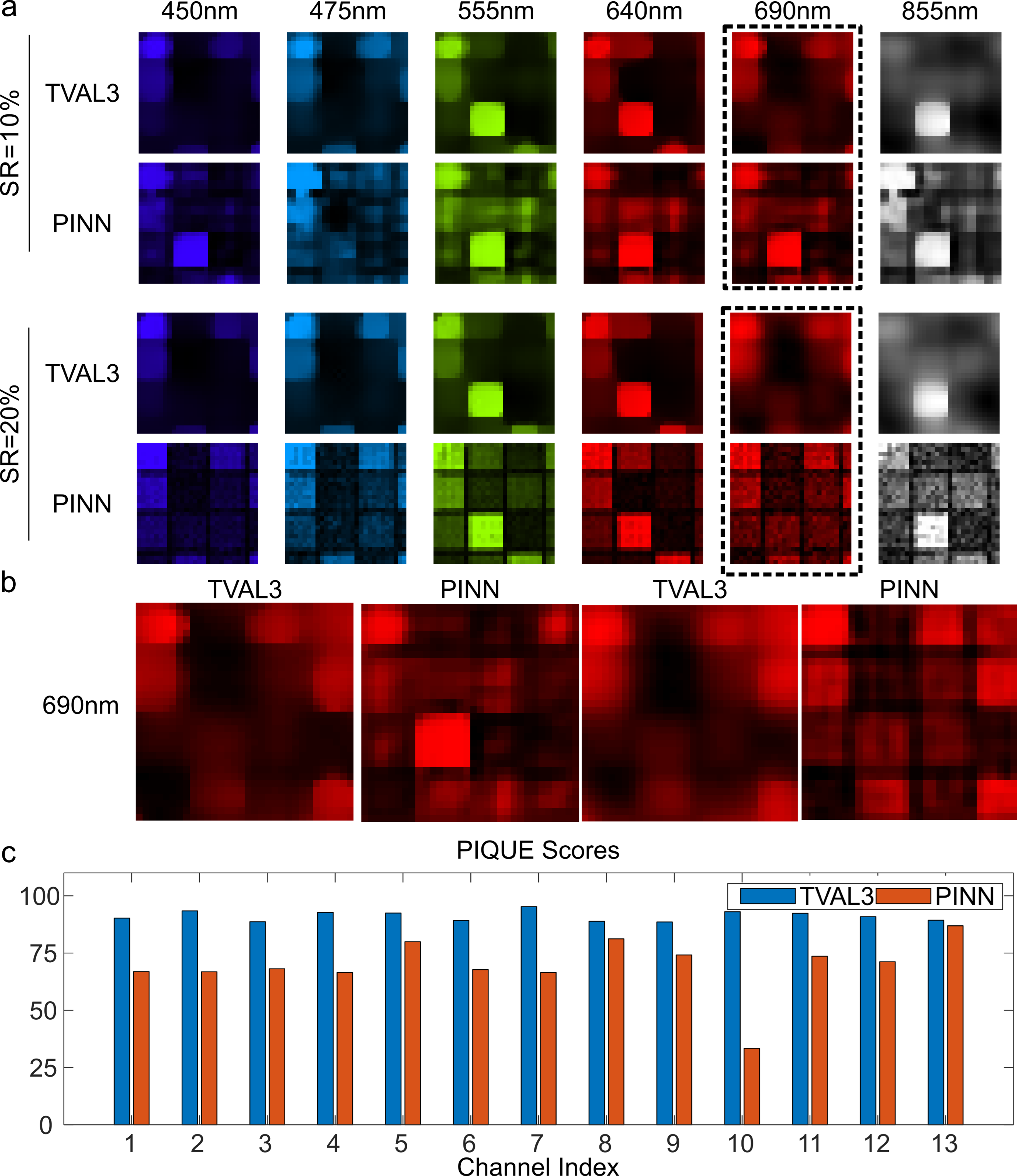}
%% Use \caption command for figure caption and label.
\caption{\justifying Characteristic of reconstructed image quality across different wavebands. (a) Reconstructed images at sampling rates of 10\% and 20\% using TVAL3 and PINN, as the wavebands of 450nm, 475nm, 555nm, 640nm, 690nm, and 855nm. (b) Zoomed-in images at 690 nm band to highlight the preservation of edge structures in PINN based reconstructions. (c) Quantified image quality with the evaluation of perception-based image quality evaluator (PIQUE) scores, showing the superior perceptual quality of PINN reconstructions.}\label{fig:4}
\end{figure*}

\begin{figure*}[htbp]%% placement specifier
    \centering%% For centre alignment of image.
    \includegraphics[width=0.8\textwidth]{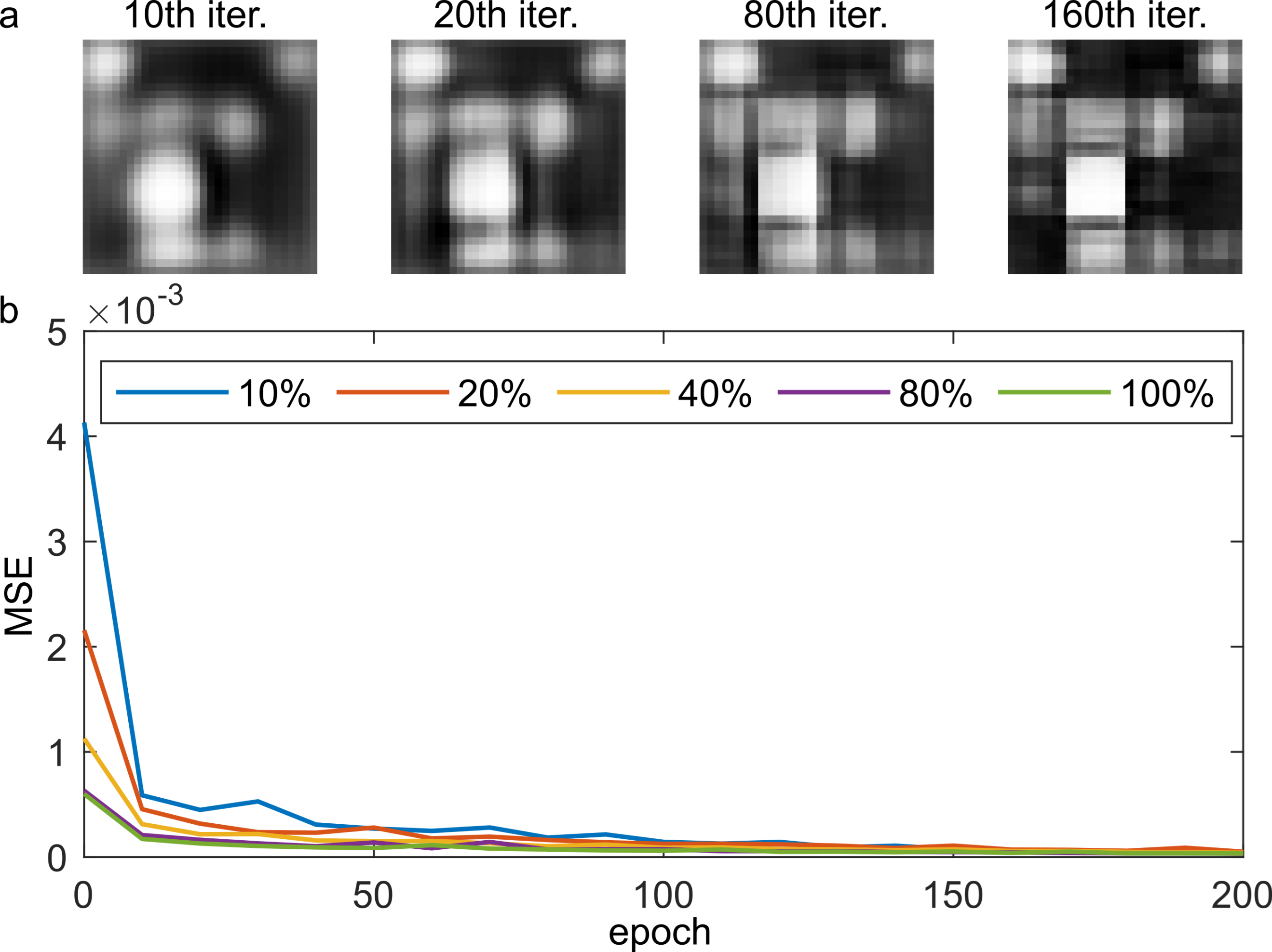}
    %% Use \caption command for figure caption and label.
    \caption{\justifying Analysis of iterative reconstruction and sampling effectiveness of PINN. (a) Reconstructed image quality under a sampling rate of 10\% in different iterations. (b) Mean square error (MSE) values between predicted measurements and raw signals under sampling rates of 10\%, 20\%, 40\%, 80\%, and 100\%.}\label{fig:5}
\end{figure*}
\begin{figure*}[tb]%% placement specifier
    \centering%% For centre alignment of image.
    \includegraphics[width=0.9\textwidth]{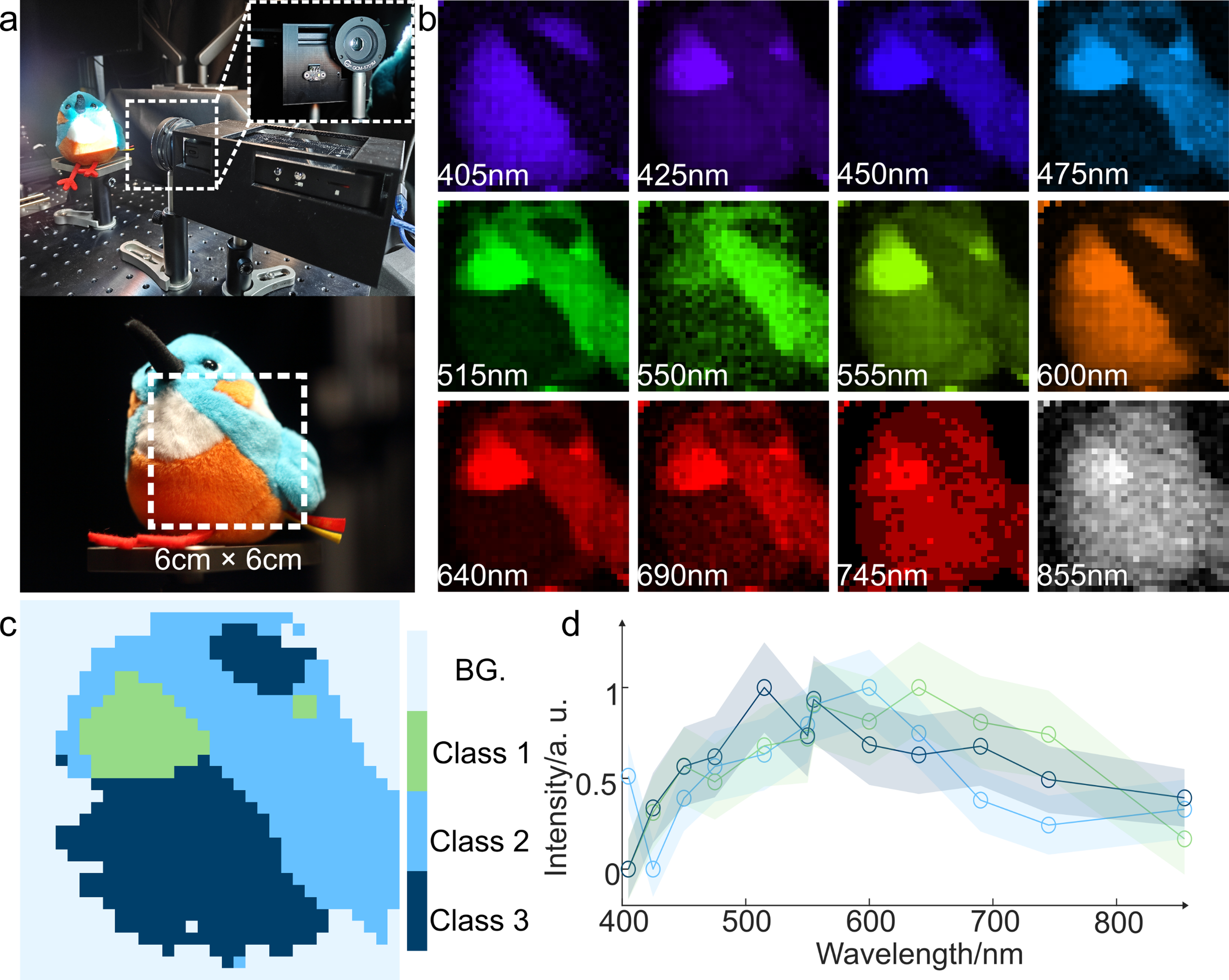}
    %% Use \caption command for figure caption and label.
    \caption{\justifying Application of the proposed MS-SPI technique in an MSI–based segmentation task. (a) The top image shows the photograph of the experimental scene, in which a plush toy is used as the object. The bottom image shows the reference RGB photograph. (b) Reconstructed multispectral images at different wavebands. (c) K-means segmentation map classified according to the multispectral images in (b). The map contains three classes with different spectral responses and the background (BG). (d) Spectral curves correspond to the spatial regions of the three classes in (c).}\label{fig:6}
\end{figure*}
To validate our method, we built a basic optical setup for multispectral imaging, as illustrated in Figure 3. Figure 3(a) shows the hardware design, in which a compact digital projector, a multispectral sensor, a data acquisition card, and an aperture are integrated to form the compact MS-SPI system. For the portability and the structural integrity, we designed a custom 3-D-printed enclosure to house all the hardware components with a size of 15.6 cm×8.7cm×7.1 cm. The illumination patterns are projected onto the surface of the target using a minimized LED projector (M9, AODIN) with a size of 14.5 cm×7.9 cm×1.8 cm, which operates at a refresh rate of 60Hz. A multi-spectral sensor (AS7343, Guangyun) is employed to measure the intensities of reflected light. The sensor’s size is approximately 3.1mm in length and 2.0mm in width, as shown in the inset of Figure 3(b). This sensor uses various optical filters positioned in front of the photodiodes to achieve distinct spectral responses.	The normalized responses are illustrated in Figure \ref{fig:2}, which were characterized by scanning a monochromatic light source with a 2 nm bandwidth and a 1 nm step size. The sensor’s integration time was set to 32 ms, corresponding to two illumination frames of the same pattern. Thus, the modulation frequency is about 30 frames per second. Each measurement was triggered immediately following the projection to ensure synchronized acquisition by utilizing a data acquisition card (USB6343, National Instruments). The following reconstruction was performed on a laptop computer.

Figure \ref{fig:4}(a) presents a comparison of image reconstruction performance using TVAL3 and PINN across multiple example spectral channels (450, 475, 555, 640, 690, and 855 nm) at two sampling rates (SRs) of 10\% and 20\%. And the image results were colored using the corresponding colors of the wavelengths. For both SRs, PINN outperforms the TVAL3 in visual fidelity and structural preservation across all examined channels (450, 475, 555, 640, 690, and 855nm). For the case with the sampling ratio of 10\%, TVAL3 achieves smooth intensity distributions but fails to recover fine spatial features, such as the edges of the blocks in the 555 nm band. In contrast, PINN reconstructs the well-defined spatial structures with higher contrast and richer texture, demonstrating more accurate and robust recovery. At the sampling rate of 20\%, although results of TVAL3 improve marginally, they remain unable to capture the details, such as dark lines between blocks in the 640 nm band, whereas PINN continues to deliver superior reconstructions with clear boundaries and enhanced textures, indicating better scalability with increased sampling. In Figure \ref{fig:4}(b), the zoomed-in images at 690 nm band show the comparison of the details.

To quantify imaging quality, we further employed the no-reference perception-based image quality evaluator (PIQUE) to assess the reconstruction fidelity at a sampling rate of 20\%, since the ground-truth reference images under identical conditions were unavailable. PIQUE estimates quality from perceptually significant spatial regions, where a smaller score indicates better perceptual quality \cite{r31}. In all spectral channels, PINN shows its advantage over TVAL3 as shown in Figure \ref{fig:4}(c). The quantitative results underscore PINN’s superiority in solving highly underdetermined inverse problems by integrating physical models within a deep-learning framework. Unlike traditional compressed sensing, which relies on handcrafted regularization, PINN leverages physical laws, achieving better generalization and higher-quality reconstructions even at low sampling rates.

Furthermore, we evaluated the performance of this neural network in terms of iterative effectiveness and physical information effectiveness. Figure \ref{fig:5}(a) presents the reconstruction outputs of 640 nm at a 10\% SR under different iteration numbers, while Figure \ref{fig:5}(b) illustrates the mean square error (MSE) curve between the physical model’s predicted single-pixel measurements and the ground truth measurements. The mentioned above together reflect the performance of the neural network. The iterative process leads to noticeable improvements in image quality, with fine details progressively revealed. Concurrently, the MSE value rapidly declines under all sampling rates, approaching convergence within 25 iterations. Integrating PINN markedly enhances measurement MSE quality across all sampling rates. In addition, reconstruction fidelity improves substantially as the sampling rate increases. As the sampling rate increases, the quality of details in the image increases significantly. This is because PINN can rely more on physical measurements to reconstruct a detailed image. It shows that PINN does not rely on the initial input, but continuously iterates to approach the physically correct result, confirming its robustness. However, at low Hadamard sampling rates, high‐frequency spatial information is irretrievably lost, leading to mosaic artifacts and preventing PINN from achieving a global optimum. This issue can be addressed by optimizing the illumination patterns and increasing the sampling rate, which in turn can further enhance the reconstruction quality.

To demonstrate the application of our method, we designed and conducted an experiment with a uniformly diffuse plush toy for the MSI–based segmentation task, as shown in Figure \ref{fig:6}. The experimental setup and the reference RGB photograph of the object are presented in Figure \ref{fig:6}(a). The multispectral reconstructed images in Figure \ref{fig:6}(b) clearly delineate the toy's shape, as the bottom-left corner labels the center wavelengths of each spectral channel. Although the imaging quality in the bands of 550 nm, 745 nm, and 855 nm is relatively poor due to low signal-to-noise ratio caused by the low illumination intensity, the strong scattering, and the small sensing area, our imaging system demonstrates overall good performance across all spectral channels. We processed the reconstructed multispectral images using an unsupervised classification algorithm, K-means \cite{r32}, to obtain pixel-wise segmentation results, which are visualized in Figure \ref{fig:6}(c). In the segmentation map, the pixels are divided into four categories, including three classes belonging to the object and the background region. The spectral curves corresponding to the three non-background categories are plotted in Figure \ref{fig:6}(d). For each spectral band, the line profiles represent the mean value across all pixels within the same category, and the shaded band indicates the standard deviation. The high accuracy achieved by the classifier indicates that the reconstructed multispectral images are qualified for image segmentation applications, and can be further enhanced by advanced classifiers and additional spectral channels.

\section{Conclusion}
In summary, we proposed a portable computational single-pixel multispectral imaging method, which employs a chip-sized spectral sensor for the compact hardware system and adapts a physics-informed untrained neural network for enhanced reconstructions. We demonstrated that utilizing a physics-informed untrained neural network in the reconstruction can markedly diminish the necessary for a large training dataset and improve the precision. The multispectral images of 12 channels with high fidelity reconstruction are acquired by using the minimized spectral sensor, instead of multiple bulky photodetectors. Moreover, we demonstrate the application of our method for an image segmentation task based on the reconstructed multispectral images.

While the proposed portable MS-SPI system demonstrates promising capabilities in compact multispectral imaging, several avenues remain for further improvement. The spectral resolution can be further improved by adopting a modified spectral sensor with more channels or reconstructing the spectral dimension data with more sophisticated algorithms \cite{r33}. The reconstruction quality could be enhanced through end-to-end joint calibration of both optical and computational components, as well as through the incorporation of physics-aware spectral unmixing strategies \cite{r34}. In addition, integrating lightweight network architectures can further accelerate the computation process \cite{r35}. Our scheme provides an alternative way to realize portable spectral sensing devices. We anticipate that this work is paving the way for future low-cost, high-resolution computational multispectral imaging solutions.

%\bigskip

\section*{Funding}
This work was supported by the National Natural Science Foundation of China (U23A20481, 62275010), the Fundamental Research Funds for the Central Universities (KG16-3549-01).
\section*{Disclosures}
The authors declare that they have no competing interests. 
\section*{Data Availability}
The data that support the findings of this study are available from the corresponding author upon reasonable request.
%\nocite{*}
\bibliographystyle{apsrev4-1}
\bibliography{ref}% Produces the bibliography via BibTeX.

\end{document}